\documentclass[conference,compsoc]{IEEEtran}
\ifCLASSOPTIONcompsoc
  \usepackage[nocompress]{cite}
\else
  \usepackage{cite}
\fi

\usepackage{hyperref}
\usepackage{url}
\usepackage{booktabs}
\usepackage{amsmath, amssymb}
\usepackage{cases}
\usepackage{textcomp}

\ifCLASSINFOpdf
  \usepackage[pdftex]{graphicx}
  \graphicspath{{./figure/}}
\else
  \usepackage[dvips]{graphicx}
  \graphicspath{{./figure/}}
  \DeclareGraphicsExtensions{.eps}
\fi
\begin{document}
%
\title{Extended Abstract: Mimicry Resilient Program Behavior Modeling \\with LSTM based Branch Models}



%
\author{\IEEEauthorblockN{
Hayoon Yi\textsuperscript{1}\IEEEauthorrefmark{1},
Gyuwan Kim\textsuperscript{1,2}\IEEEauthorrefmark{1}, 
Jangho Lee\textsuperscript{1}, 
Sunwoo Ahn\textsuperscript{1}, 
Younghan Lee\textsuperscript{1}, 
Sungroh Yoon\textsuperscript{1}\IEEEauthorrefmark{2}, 
Yunheung Paek\textsuperscript{1}\IEEEauthorrefmark{2}}
\IEEEauthorblockA{
\textsuperscript{1}Dept. of Electrical and Computer Engineering, Seoul National University\\
\textsuperscript{2}Search Solutions, Inc\\
Email: {hyyi,kgwmath,ubuntu,swahn,yhlee,sryoon,ypaek}@snu.ac.kr\\
\IEEEauthorrefmark{1}: Equal Contribution, 
\IEEEauthorrefmark{2}: Corresponding Author
}
}


\maketitle


\begin{abstract}

In the software design, protecting a computer system from a plethora of software attacks or malware in the wild has been increasingly important. One branch of research to detect the existence of attacks or malware, there has been much work focused on modeling the runtime behavior of a program. Stemming from the seminal work of Forrest et al., one of the main tools to model program behavior is system call sequences. Unfortunately, however, since mimicry attacks were proposed, program behavior models based solely on system call sequences could no longer ensure the security of systems and require additional information that comes with its own drawbacks. In this paper, we report our preliminary findings in our research to build a mimicry resilient program behavior model that has lesser drawbacks. We employ branch sequences to harden our program behavior model against mimicry attacks while employing hardware features for efficient extraction of such branch information during program runtime. In order to handle the large scale of branch sequences, we also employ LSTM, the de facto standard in deep learning based sequence modeling and report our preliminary experiments on its interaction with program branch sequences.

\end{abstract}


%
\IEEEpeerreviewmaketitle

\section{Introduction}
\label{sec:intro}
Protecting a computer system from a plethora of software attacks or malware in the wild has been increasingly important. Although the nature or cause of an attack is often hard to know in practice, it usually results in anomalous behavior different from what can be seen in a normal program during execution. The rationale behind this argument is that in the case of attacks, to infiltrate a system, attackers usually gain control over program execution exploiting exposed vulnerabilities, which resultantly produces different program behavior from that of benign programs. Following this logic, as one branch of research to detect the existence of attacks or malware, there has been much work focused on modeling the runtime behavior of a program~\cite{hofmeyr1998, hoang2003, hu2009, yolacan2014,shu2015unearthing, xu2016sharper, staudemeyer2013, staudemeyer2015applying, pascanu2015malware, kolosnjaji2016deep}. This is done by either modeling the behavior of normal program execution in order to detect attacks that cause anomalies or modeling the behavior of known malware families to detect similar malware. 

Stemming from the seminal work of Forrest et al.~\cite{forrest1996sense}, one of the main tools to model program behavior is system call sequences. As stated in ~\cite{forrest1996sense}, for most malware or attacks to function correctly, they must access system resources which are only accessible through system calls to the OS. Therefore, we could model the system call sequences representing such accesses in order to discern malicious activity or we could model the expected normal system call sequences of a system and discern any anomalous sequence that does not follow the model to be potentially malicious.

Unfortunately, since \textit{mimicry} attacks~\cite{wagner2002mimicry, parampalli2008practical}, which hide malicious system call sequences by mimicking that of benign programs, were proposed, program behavior models based solely on system call sequences could no longer ensure the security of systems. In order to counter mimicry attacks, researchers aimed to include additional information, such as system call arguments~\cite{mutz2006anomalous,maggi2010detecting} or call stack information~\cite{feng2003anomaly}, that could help build a program behavior model capable of differentiating true normal system call sequences from mimicked system call sequences. However, the inclusion of additional information brings its own drawbacks. For instance, unlike system call sequences that could be easily modeled automatically via various sequence based machine learning methods, system call arguments come in many different forms and require complex handcrafted constraints or augmentations to model every different type of argument. While this makes it capable of leveraging human intuition, it also leaves it susceptible to human error. 

Therefore, in this paper, we report our preliminary findings in our research to build a mimicry resilient program behavior model that does not suffer from the drawbacks of prior work. As, during our preliminary studies, we have found that no-op system calls~\cite{wagner2002mimicry}, which are the main tool of building mimicry attacks, are easily discernible with the help of branch traces, We employ branch sequences to harden our program behavior model against mimicry attacks. With the help of recent hardware features, Intel Processor Tracing (PT)~\cite{pt} on Intel x64 architectures and real-time trace macrocells (ETM, PTM, STM) in ARM architectures, we can acquire branch sequences of a running program with low overhead. Furthermore, branch traces magnify any anomalies in a corresponding system call trace as an anomalous system call would bring with it many anomalous branch instructions. However, by employing branch traces, the possible number of candidates for each element in a sequence are several orders of magnitude larger than that of system call sequences. Due to this fact, most system call sequence modeling methods proposed in prior work would not work well with branch traces. In order to address this, we leverage deep learning, more specifically Long Short Term Memory (LSTM)~\cite{hochreiter1997}, to handle large scale sequence modeling. LSTM shows great success in various sequence modeling applications and could be considered the current de facto standard for deep learning based sequence modeling. Our preliminary experiments show it has great promise in branch sequence modeling. Based on these, we design a prototype system, DeePBM which leverages Intel PT to acquire and deliver branch sequences for the training and inference of an LSTM based program behavior model. 

The rest of this paper is organized as follows. First we give a brief overview of the design of our prototype system DeePBM and our LSTM branch model in Section~\ref{sec:design}. Then we report our preliminary findings in modeling program branch sequences with LSTM in Section~\ref{sec:findings}. Finally, we conclude our paper in Section~\ref{sec:conclusion}.

\section{Prototype Design}
\label{sec:design}

\begin{figure}[t]
         	\centering
         	\includegraphics[width=1.0\columnwidth]{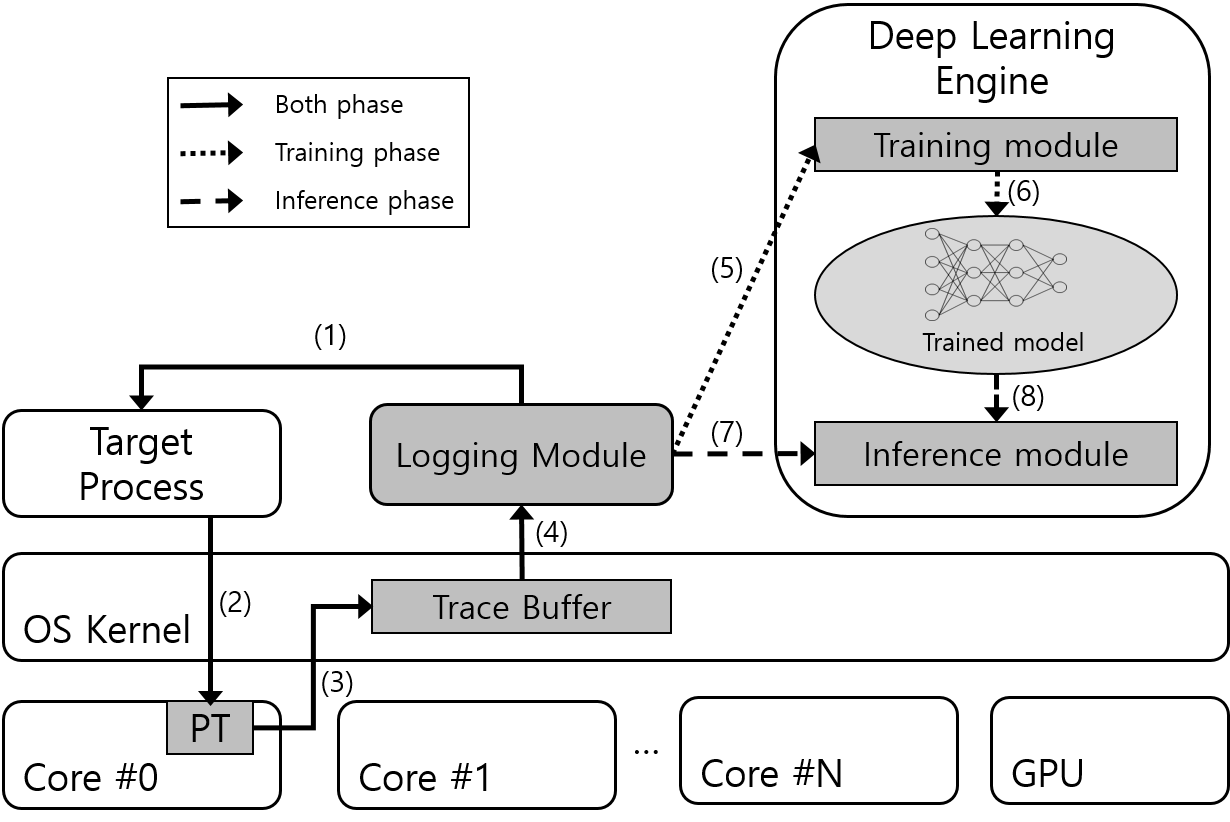}
            \vspace*{-10pt}
            \caption{Architectural overview of the DeePBM framework.}
         	\label{fig:design_overview}
\vspace*{-15pt}
\end{figure}

The ultimate goal of DeePBM is to support an LSTM branch model with an efficient runtime mechanism that can record all the branch outcomes produced during program execution and deliver them to the model. In this section, we describe the overall architecture of DeePBM. 

\subsection{Components of DeePBM}

Logging module (LM) keeps track of execution paths of the live target process, gathering through PT its branch traces generated by the CPU. LM aims to capture the complete behavior of the target process by collecting branch traces from both user mode and kernel mode. In addition, if the process forks a child process during execution, LM will be immediately configured  to  monitor it as well.

As a hardware extension to support control flow tracing, PT generates branch traces in the form of several encoded data packets including Taken Not-Taken (TNT) packets, Target IP (TIP) packets, and Flow Update packets (FUP). TNT indicates the direction of direct conditional branches and represent the information about returns. TIP records the target address of control transfers such as indirect jumps and indirect calls. FUP provides the source address for asynchronous events such as those triggering interrupts. In the current implementation, LM gathers the traces of branch target addresses from TNT and TIP packets. 

Deep learning engine (DLE) consists of two modules each for training and inference, respectively. Under a controlled environment for learning program behavior, the training module performs four types of operations belonging to the training phase of our deep learning model. One operation is dividing available branch traces either into the training set or validation set. Another is transforming branch traces into an input-friendly form for model training, which will be described in Section~\ref{sec:dle}. The remaining two are training the model with the training set and validating the model with the validation set. In our implementation, these operations are repeated as many times as needed. The inference module performs two types of operations belonging to the inference phase of the model. One is transforming the given branch trace into an input-friendly form for inference. The other is detecting anomalies or patterns in the given branch trace with the trained branch model and reporting the result. In Section~\ref{sec:dle}, we will discuss the deep learning model trained and managed by DLE.

\begin{figure}[t]
\centering
\includegraphics[width=\columnwidth]{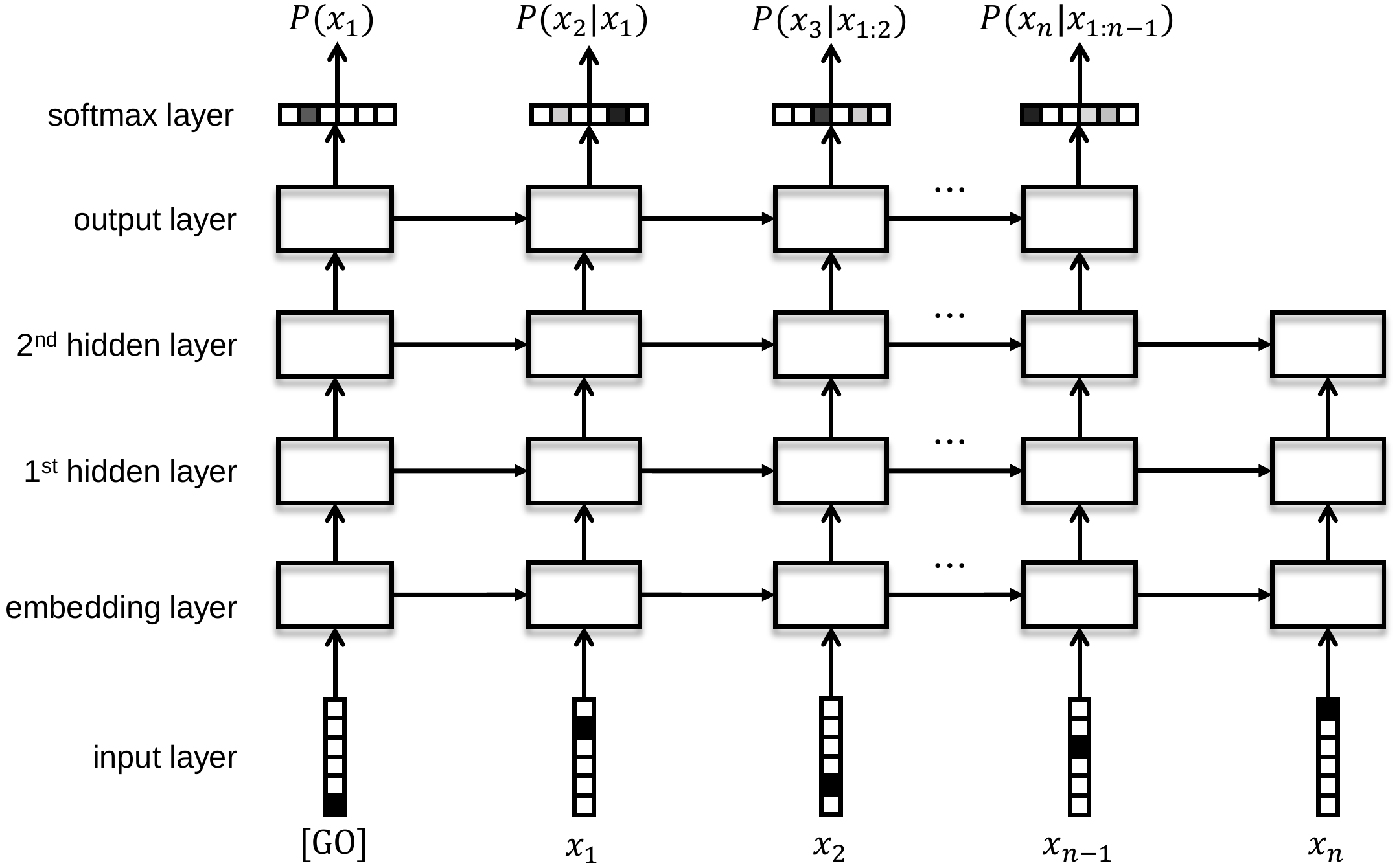}
 \vspace*{-10pt}
\caption{Branch model}
\vspace*{-15pt}
\label{fig:architecture}
\end{figure}

\subsection{LSTM branch model}
\label{sec:dle}
As typical processes in modern programs execute long chains of branches, the number of branches required to fully understand the meaning of program behavior is quite large. In addition, the branches comprising a process are intertwined with each other in a complicated way. In this regard, learning long-term dependence is crucial for devising effective program behavior models. Therefore, we employ long short term memory (LSTM)~\cite{hochreiter1997}, a well-designed RNN architecture component that has the capability of capturing long term dependencies, as the basis of our DLE. 

Figure~\ref{fig:architecture} illustrates the architecture of the \textit {branch model}, that is, our language model of branch sequences, which estimates the probability distribution of the next branch in a sequence given the sequence of previous branch events. Let vocabulary $V$ be the set of all possible branch target addresses. Then, each branch target address is indexed by an integer starting from $1$ to $K=|V|$. However, in general, it is hard to know all possible branch target addresses beforehand and they are not fixed. To deal with this issue, we build a vocabulary that consists of the top $K-1$ most frequent addresses and a single special address, \textit{unknown}, which represents all the other addresses. Let $x=x_1x_2\cdots x_l (x_i\in V)$ denote a sequence of $l$ branches.

At the input layer, the branch at each time step $x_i$ is fed into the model in the form of one-hot encoding, in other words, a $K$ dimensional vector with all elements zero except position $x_i$. At the embedding layer, incoming branches are embedded to continuous space by multiplying embedding matrix $W$, which should be learned. At the hidden layer, the LSTM unit has an internal state, and this state is updated recurrently at each time step. At the output layer, a softmax activation function is used to produce the estimation of normalized probability values of possible branches coming next in the sequence, $P(x_i|x_{1:i-1})$. According to the chain rule, we can estimate the sequence probability by the following formula:
\begin{align}
P(x)=\prod_{i=1}^{l}P(x_i|x_{1:i-1})
\end{align}

Given training branch sequence data, we can train this LSTM-based branch model using the backpropagation through time (BPTT) algorithm. The training criterion minimizes the cross-entropy loss, which is equivalent to maximizing the likelihood of the branch sequence. A standard RNN often suffers from the vanishing/exploding gradient problem, and when training with BPTT, gradient values tend to blow up or vanish exponentially. This makes it difficult to learn long-term dependency in RNNs~\cite{bengio1994}. LSTM is equipped with an explicit memory cell and tends to be more effective to cope with this problem. Given a new query branch sequence, on the assumption that attack/malware branch patterns deviate from normal patterns, a sequence with a perplexity, an average negative log-likelihood probability, above a threshold could be considered to not be likely of what is learned by the model. In other words, a sequence that shows high perplexity in a branch model of a malware family is probably not associated with that particular malware type.

\section{Preliminary findings}
\label{sec:findings}
In this section, we share our findings during our preliminary studies and experiments. We first explore the inherent resilience branch sequences offer against mimicry attacks. Then we share our experiments in employing DeePBM to learn the normal behaviors of programs in order to detect any attacks against those programs.

\begin{figure}[t]
	\centering
	\includegraphics[width=\columnwidth]{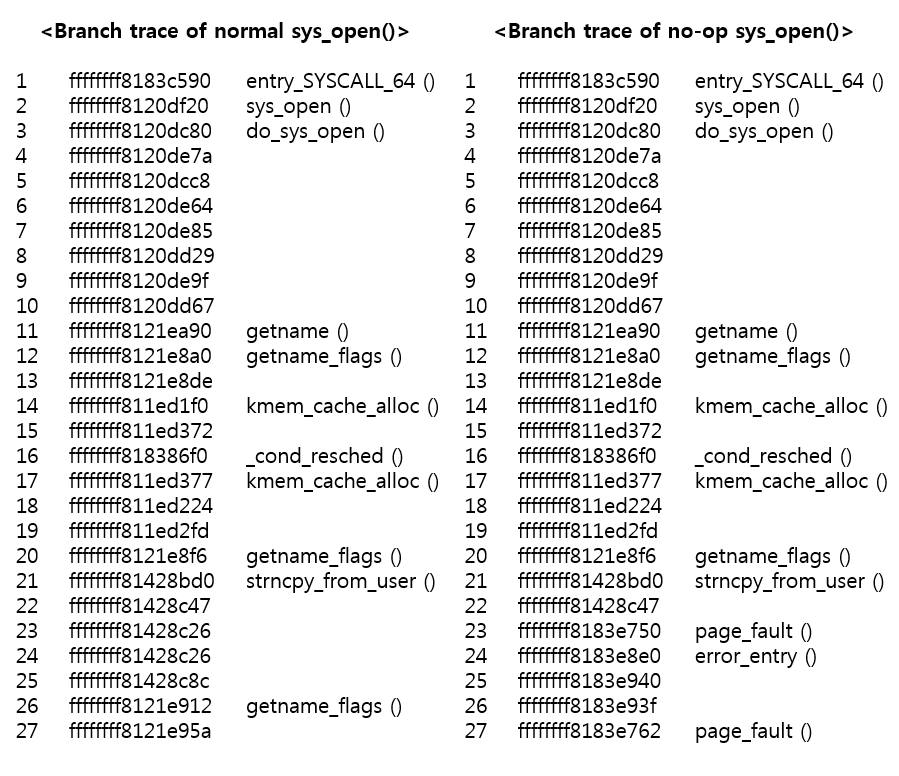}
     \vspace*{-20pt}
	\caption{Initial branch sequence of system call open()}
           \vspace*{-15pt}
	\label{fig:sys_open}
\end{figure}

\begin{figure*}[t]
\centering
\includegraphics[width=1.\textwidth]{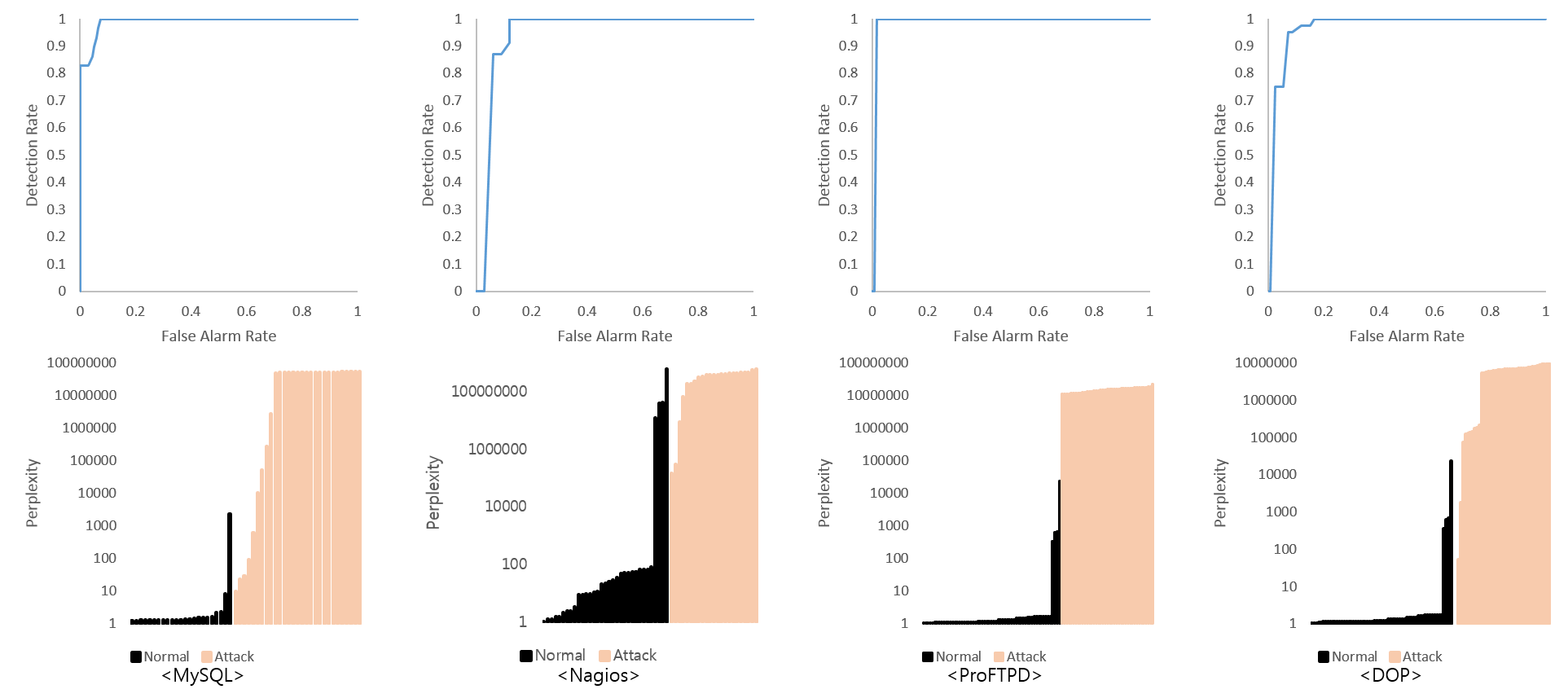}
           \vspace*{-20pt}
\caption{The ROC curves and perplexity of each Program. ProFTPD and DOP share the same normal sequence in this figure.}
\label{fig:perplexity}
           \vspace*{-15pt}
\end{figure*}

\subsection{Branch sequences and Mimicry attacks}
\textit{Mimicry} attacks~\cite{wagner2002mimicry, parampalli2008practical} were originally proposed to defeat solutions that only rely on system call sequences. In~\cite{wagner2002mimicry}, a handcrafted mimicry attack sequence is given by transforming an existing attack to emit a system call sequence that would be viewed as a normal sequence by existing work. The transformation relies on no-op system calls which can be made in such a way that it would not affect anything, such as calling mkdir() with an invalid pointer, to hide the system call footprints of the malicious code. This technique would nullify program behavior modeling solutions that only monitor system calls, and subsequently necessitates supplementary data in order to differentiate no-op system calls between normal system calls. During our preliminary experimentation we have found that by simply examining branch sequences, which would be counted as using supplementary data in addition to system calls, we were able to isolate no-op calls from normal ones when the branch sequences following system calls diverge even though the system calls are of the same type. For example, Figure~\ref{fig:sys_open} shows the initial 27 branches when an open() system call is made. As can be seen, the sequence of a normal call and that of a no-op call diverges from the 23rd branch. After the shown branches, the normal system call goes through 364 branches handling opening the file, while the no-op call goes through 159 branches handling errors stemming from trying to use a null pointer. This trend can be seen through most no-op system calls, as they leverage null or invalid pointers to render system calls void. Therefore, we would be able to discover system call sequence mimicry attacks in past literature. However theoretically, any program behavior modeling solution including ours is susceptible to new mimicry attacks crafted to forge mimic sequence patterns similar to those of call/branch sequences accepted to be normal by its existing model~\cite{sharif2007understanding}. The key issue is how much leeway is left for adversaries to maneuver. Solutions that examine more detailed information leave less room for new mimicry attacks. This is another benefit of operating on branch sequences, that is, the most detailed information which is efficiently available. Though theoretically there still might exist mimicry attacks that would not be detectable by branch models, the expressiveness of such attacks would be severely limited as it will be hard to find necessary no-op branches. Therefore, we believe in practice that it would be considerably difficult to build mimicry attacks to avoid detection from program branch models.

\subsection{Branch sequence model for anomaly detection}
We trained branch models on three separate real programs: MySQL, Nagios and ProFTPD~\cite{mysql, nagios, proftpd}. These applications were mainly selected as they were the target victims of four publicly available attacks: privilege escalation via CVE-2016-6663 and CVE-2016-6664~\cite{mysql_attack1, mysql_attack2} against MySQL, privilege escalation via CVE-2016-9565 and CVE-2016-9566~\cite{nagios_attack1, nagios_attack2} against Nagios and address leakage attack and data-oriented programming (DOP) attack via CVE-2006-5815~\cite{hu2016data, dop_attack}.

The lower row of Figure~\ref{fig:perplexity} depicts the ROC curves and perplexity of sequences collected for inference. This shows how our branch model regards the collected sequences. As can be seen, the model is quite capable of discerning between the normal and attack sequences. The perplexity for the normal sequences are quite low, meaning that they conform to what our model has learned from the training datasets. On the other hand, nearly all sequences containing attacks show high perplexity, which would mean the branch sequences deviate from what our model expected. For a few normal sequences, the model reports high perplexity. With further examination, we believe this is due to the sequences containing rare events, such as one of the Nagios monitored computers crashing, that were not covered by the training data. This could also be seen that our model can successfully interpret branch sequences of such events as an anomaly in program behavior. As seen in the upper row of Figure~\ref{fig:perplexity}, the overall detection performance of DeePBM's trained models are reflected in their ROC curves. The models can achieve high detection rate with low false positive rates on modern server programs.

\section{Conclusion}
\label{sec:conclusion}
In this paper, we have shared our preliminary findings on applying LSTM to branch sequences in order to model program behavior. We believe our work has provided a glimpse of evidence that can demonstrate the feasibility of deep learning models employed for program behavior modeling via LSTM and branch sequences. We have also briefly explored the possibility of mimicry attack mitigation through the nature of branch sequences and believe when leveraged well, branch information could provide an efficient way of mimicry mitigation. 

As an extension of our work, we are currently looking into modeling malware families through branch sequences and testing its viability in finding similar mutated malware samples as well as testing our work against actual working mimicry malware.

\section{Acknowledgements}
This work was supported in part by the National Research Foundation of Korea (NRF) grant funded by the Korea government (MSIT) [2018R1A2B3001628] and [NRF-2017R1A2A1A17069478], in part by the MSIT, Korea under the ITRC support program(IITP-2017-2015-0-00403) supervised by the IITP and in part by the IITP grant funded by the Korea government(MSIT) (2015-0-00573).






%
\bibliographystyle{IEEEtranS}
\bibliography{padon}

\end{document}